\documentclass[twocolumn,prl,aps,groupedaddress,superscriptaddress]{revtex4}
\usepackage[latin9]{inputenc}
\usepackage{color}
\usepackage{amstext}
\usepackage{amssymb}
\usepackage{graphicx}
\usepackage{dsfont}
\usepackage[unicode=true,pdfusetitle,
 bookmarks=false,
 breaklinks=false,pdfborder={0 0 1},backref=false,colorlinks=true]
 {hyperref}
\hypersetup{
 bookmarksnumbered=false,bookmarksopen=false,citecolor=blue,linkcolor=blue}

\usepackage{amsmath}
\usepackage{txfonts}
\usepackage{textcomp}
\usepackage{tikz}
\usetikzlibrary{arrows}
\usetikzlibrary{shapes}
\usetikzlibrary{fit}
\usetikzlibrary{positioning}

\usepackage[caption=false]{subfig}

\newcommand{\T}{\Theta}
\newcommand{\X}{\mathcal{X}}
\newcommand{\p}{\mathcal{P}}

\newcommand{\Carb}{$^{13}$C }
\newcommand{\Nitro}{$^{14}$N }

\newcommand{\ket}[1]{\left|{#1}\right>}
\newcommand{\bra}[1]{\left<{#1}\right|}
\newcommand{\ketbra}[2]{\left|{#1}\right>\left<{#2}\right|}

\newcommand{\tr}[1]{\mathrm{tr}\left[{#1}\right]}
\newcommand{\ptr}[2]{\mathrm{tr}_{#1}\left[{#2}\right]}

\makeatletter

\@ifundefined{textcolor}{}
{%
 \definecolor{BLACK}{gray}{0}
 \definecolor{WHITE}{gray}{1}
 \definecolor{RED}{rgb}{1,0,0}
 \definecolor{GREEN}{rgb}{0,0.5,0}
 \definecolor{BLUE}{rgb}{0,0,1}
 \definecolor{CYAN}{cmyk}{1,0,0,0}
 \definecolor{MAGENTA}{cmyk}{0,1,0,0}
 \definecolor{YELLOW}{cmyk}{0,0,1,0}
}

\makeatother

\begin{document}

\title{Controllable Non-Markovianity for a Spin Qubit in Diamond}

\author{J. F. Haase}
\affiliation{Institut f\"ur Theoretische Physik and IQST, Albert-Einstein-Allee 11, Universit\"at Ulm, D-89069 Ulm, Germany}
\author{P. J. Vetter}
\author{T. Unden}
\affiliation{Institut f\"ur Quantenoptik and IQST, Albert-Einstein-Allee 11, Universit\"at Ulm, D-89069 Ulm, Germany}
\author{A. Smirne}
\author{J. Rosskopf}
\affiliation{Institut f\"ur Theoretische Physik and IQST, Albert-Einstein-Allee 11, Universit\"at Ulm, D-89069 Ulm, Germany}
\author{\mbox{B. Naydenov}}
\affiliation{Institut f\"ur Quantenoptik and IQST, Albert-Einstein-Allee 11, Universit\"at Ulm, D-89069 Ulm, Germany}
\author{A. Stacey}
\affiliation{Element Six, Harwell Campus, Fermi Avenue, Didcot OX11 0QR, United Kingdom}
\affiliation{Centre for Quantum Computation and Communication Technology, School of Physics, University of Melbourne, Parkville, Melbourne VIC 3010, Australia}
\author{F. Jelezko}
\affiliation{Institut f\"ur Quantenoptik and IQST, Albert-Einstein-Allee 11, Universit\"at Ulm, D-89069 Ulm, Germany}
\author{M. B. Plenio}
\author{S. F. Huelga}
\affiliation{Institut f\"ur Theoretische Physik and IQST, Albert-Einstein-Allee 11, Universit\"at Ulm, D-89069 Ulm, Germany}

\begin{abstract}
We present a flexible scheme to realize non-Markovian dynamics of an electronic spin qubit, using a nitrogen-vacancy center in diamond where the inherent nitrogen spin serves as a regulator of the dynamics. By changing the population of the nitrogen spin, we show that we can smoothly tune the non-Markovianity of the electron spin's dynamics. Furthermore, we examine the decoherence dynamics induced by the spin bath to exclude other sources of non-Markovianity. The amount of collected measurement data is kept at a minimum by employing Bayesian data analysis. This allows for a precise quantification of the parameters involved in the description of the dynamics and a prediction of so far unobserved data points.
\end{abstract}

\maketitle

\textit{Introduction.---}
Realistic physical systems are subject to environmental noise which affects their quantum dynamics \cite{Carmichael1993,Breuer2002,Gardiner2004,Rivas2012}.
The rapidly advancing development of quantum technologies which are aiming to make us of quantum dynamics in a broad range of applications such as quantum computing \cite{Nielsen2000}, quantum cryptography \cite{Gisin2002}, quantum simulation \cite{Georgescu2014}, quantum sensing \cite{Degen2017} and quantum metrology \cite{Giovannetti2004} calls for a detailed understanding of these noise sources that may alter their function.

Typically, environmental noise does not induce featureless white noise on the system, but can exhibit spatial and temporal correlations that can be used when addressing the system-environment interaction. Non-Markovian noise, that is the subject of this work, exhibits temporal correlation originating from some slow internal evolution of the environment \cite{Breuer2002,Rivas2014,Breuer2016,deVega2017}. On the one hand, one may combat such non-Markovian noise by means of dynamical decoupling methods,
which allow to partially shield the system of interest from the impact of noise \cite{Cywinski2008, Ryan2010, Cai2012}.
On the other hand, it has been recognised early on that noise may also be a resource, e.g., for the generation of entangled states \cite{Plenio2008, Huelga2012}. In particular, one may explore the specific advantages that colored noise can provide here; this has been shown in several reports \cite{Vasile2011,Schmidt2011,Huelga2012,Laine2014,Bylicka2014,Chin2012,Chin2013,Dong2018,Torre2018}.
More recently, the introduction of definite and general ways to 
quantify the degree of non-Markovianity of quantum dynamics
\cite{Wolf2008,Breuer2009,Rivas2010,Lorenzo2013,Chruscinski2014,Rivas2014,Torre2015,Breuer2016}
has provided a further boost for the quantitative understanding of the role of non-Markovianity
in different settings and has increased the ability to manipulate open-system dynamics,
in view of possible strategies to reduce the detrimental effects of noise. In fact, an extended control over the amount of non-Markovianity has been demonstrated experimentally
in trapped ion systems \cite{Wittemer2018} and photonic setups \cite{Liu2011,Cialdi2011,Chiuri2012,Jin2015}.

Here, we want to take a further step in the direction of the full control of the non-Markovianity of 
quantum dynamics, by investigating theoretically and experimentally
the different dynamical regimes experienced by an electronic spin qubit of a 
nitrogen-vacancy center (NV) in diamond \cite{Manson06, Doherty13}. We stress that 
the system at hand is undergoing a genuine open-system evolution,
in which the main source of noise 
inducing non-Markovianity, namely, the nitrogen nuclear spin is inherent part of the NV center.
The procedure in our work consists of two steps: First a characterization of the natural background noise to exclude any source of non-Markovianity besides the nitrogen spin. Therefore we examine the free-induction decay (FID) of the electron spin while the interaction with the nitrogen spin is suppressed. The FID is induced by various sources, such as \Carb spins or additional nitrogen impurities, the diamond surface, but also experimental limitations, e.g., drifts in the optical setup. We show that the obtained data can be analyzed efficiently using Bayesian inference methods \cite{Sivia2006, Kruschke2015, Sharma2017, Schwartz2017}. These allow for a large number of free parameters and determine from a multi-dimensional probability distribution the most likely parameter set describing the data. They are therefore particularly well-suited to fully characterize the open-system dynamics at hand. 
Secondly, we study how to use the nitrogen spin inherent to the NV center to control the degree of non-Markovianity of the electronic spin. Therefore, we manipulate the polarization of the nitrogen spin to induce collapses and revivals on the electronic spin coherence, while the polarization direction of the nitrogen spin defines the amplitude of these collapses and revivals. The degree of non-Markovianity corresponding to the different configurations is measured and compared 
with the theoretical predictions provided by the Bayesian data analysis,
showing that we can achieve a full control on the amount of non-Markovianity involved
in the evolution of this solid-state system.  
\\

\begin{figure}[t!]
\begin{centering}
\vspace{0cm}
 \hspace{-0.3cm} \includegraphics[width=0.95\columnwidth]{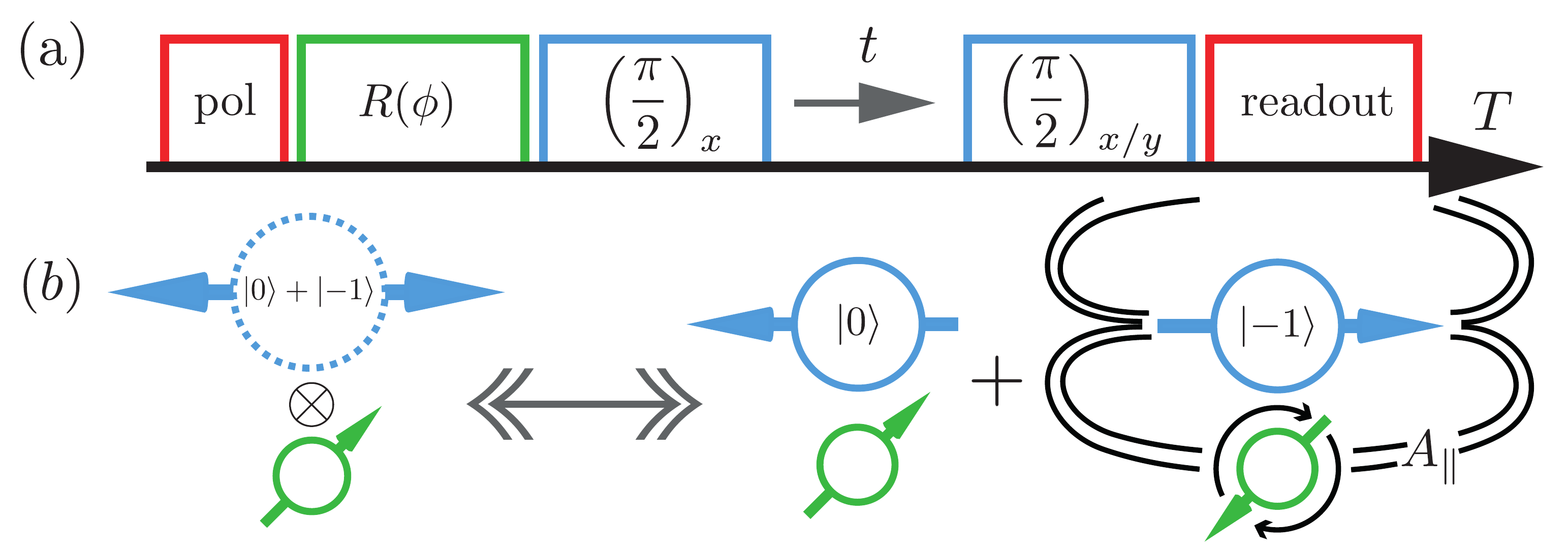} 
\par\end{centering}
\caption{(Color online) The sequence for the Ramsey experiments in (a) consists of a preparation phase, where the pulse $R(\phi)$ can control the population of the \Nitro spin, two $\pi/2$ pulses (either $x$ or $y$ phase) on the electron spin and the subsequent readout. During the free period $t$, the spins undergo the conditional evolution illustrated in (b). If the electron spin populates $\ket{-1}$, it produces a hyperfine field which induces rotations of the \Nitro spin. Therefore the pair switches continuously between a product (left) and an entangled state (right) [the roles of \Nitro and electron spin are interchangeable].}
\label{fig:FigTheory}
\end{figure}

\textit{Model.---}
The NV center is a point defect in the diamond lattice consisting of a substitutional nitrogen atom adjacent to a vacancy. Its negatively charged state possesses an electronic spin triplet ${}^{3}\text{A}$ ground state \cite{Manson06} with a zero field splitting of $\Delta=2\pi\cdot2.87~\text{GHz}$ between the $\ket{m_s=0}$ and $\ket{m_s=\pm1}$ states (from now on we denote $S_z\ket{m_s=i}=i\ket{m_s=i}=i\ket{i}$). Interaction with the inherent nitrogen nuclear spin results in a hyperfine splitting of the $\ket{\pm1}$ states, depending on the nitrogen isotope, here $^{14}$N ($I=1$), which results in a hyperfine splitting of $A_{\parallel}\approx2.14~\text{MHz}$ \cite{Felton09}. We use a low nitrogen ($<1\,\mathrm{ppb}$) diamond with a concentration of 0.2\% \Carb nuclear spins to prolong the electron spin coherence time. We identified a native NV center, located deep (few \textmu m) below the diamond surface.
The Hamiltonian of this configuration is given by \cite{Doherty13}
\begin{eqnarray}
H_{\mathrm{lab}} &=& \Delta S_z^2 + \gamma_e B_z S_z + P I_z^2 +\gamma_N B_z I_z \notag \\ 
&&+ S_z A_{\parallel} I_z + A_{\perp} \left(S_xI_x + S_yI_y\right) +  H_R
\end{eqnarray}
where $S (I)$ are the electron (\Nitro) spin-1 operators, $B_z$ a magnetic field applied along the NV symmetry axis and the electronic (\Nitro) gyromagnetic ratio is labeled by $\gamma_e$ ($\gamma_N$), the quadrupole splitting $P$ and orthogonal interaction $A_\perp$. An applied field of $B_z=453\,$G lifts the degeneracy between the $\ket{\pm1}$ states. The Hamiltonian $H_R$ contains all remaining terms originating from the environment of the NV, e.g. \Carb spins and other nitrogen impurities, including their coupling to the electron spin, but may also considered as an effective Hamiltonian responsible for experimental imperfections \cite{Maze2012,Romach2015}. We apply the secular approximation due to the large zero field splitting $\Delta \gg A_\perp\approx 2\pi \cdot2.70\, \mathrm{MHz}$ \cite{Felton09}, which prohibits flips of the \Nitro spin and also removes all terms in $H_R$ not coupling to $S_z$ \cite{Maze2012}. Because all free energy terms commute with the remaining interaction Hamiltonian $S_z A_\parallel I_z$, these terms can be removed in a rotating frame yielding
\begin{equation}
H = S_z A_\parallel I_z + H_R.
\label{eq:rotHam}
\end{equation}  

We employ the electron spin as a noise sensor for the environment choosing the subspace spanned by the $\ket{0}$ and $\ket{-1}$ state as an artificial qubit. Because of the pure dephasing Hamiltonian, the reduced density matrix of the electron spin only experiences a modulation of the coherence elements, hence the FID is efficiently measured by a Ramsey experiment, whose scheme is sketched in Fig.\ref{fig:FigTheory}(a). The electron spin preparation and readout is achieved optically.
Spin-selective, non-radiative inter-system crossing to a metastable singlet state between electronic excited and ground state \cite{Manson06} enables a strong electron spin polarization into the $\ket{0}$ ground state. The higher photoluminescence intensity of the $\ket{0}$ state allows to determine the electron spin state. We polarize the nitrogen nuclear spin in the $\ket{m_I=1}$ state by optical pumping \cite{Jacques09} and rotate it by a radio frequency pulse $R(\phi)$ to a desired coherent state.
After polarization, a $\pi/2$ pulse flips the electron spin to the superposition state $\ket{\psi}=(\ket{0}+\ket{-1})/\sqrt{2}$. For a time $t$ the system will evolve freely depending on the electron spin state as depicted in Fig.\ref{fig:FigTheory}(b), i.e. according to the conditional Hamiltonian $H_i=\bra{i}H\ket{i}$. Assuming an initial product state, $\rho = \rho^{(e)} \otimes \rho^{(N)} \otimes \rho^{(R)}$ (with $\rho^{(e)}=\ketbra{\psi}{\psi}$ and $\rho^{(R)}$ arbitrary), the dynamic of the electron spin is completely described by the coherence modulation, i.e.
\begin{equation}
\rho_{0,-1}^{(e)}(t) \propto \bra{0}\ptr{N,R}{\rho(t)}\ket{-1} = \ptr{N,R}{e^{-itH_0}\rho^{(N)}\otimes\rho^{(R)}e^{itH_{-1}}},
\end{equation}
where $\ptr{N,R}{\bullet}$ denotes the partial trace over the nitrogen and bath degrees of freedom. Assuming no residual population left in $\ket{-1}$, the length of the Bloch vector associated with the qubit in the $\lbrace \ket{0},\ket{1}\rbrace$ subspace is equivalent to the coherence. This length can directly be calculated as 
\begin{eqnarray}
r(t) = &&\Big[p_0^2+p_1^2+p_{-1}^2+2p_0(p_1+p_{-1})\cos(A_{\parallel}t)  \nonumber \\
&& +2p_1p_{-1}\cos(2A_{\parallel}t) \Big] ^{1/2}\,|L(t)|,
\label{eq:BlochLength}
\end{eqnarray}
where $L(t) = \tr{e^{-it\bra{0}H_R\ket{0}}\rho^{(R)}e^{it\bra{-1}H_R\ket{-1}}}$ and $p_i$ is the initial population in the state $\ket{m_I=i}$ of the nitrogen spin. Using the normalization constraint, we parameterize $p_1=p \cos^2(\phi/2)$, $p_0=p \sin^2(\phi/2)$ and $p_{-1}=1-p$ where $\phi$ is a mixing angle and $p$ the amount of population in the desired subspace of $\ket{m_I = 0,1}$. For the readout, the electron spin is rotated back to the z-axis (either around $x$ or $y$) and after a subsequent readout pulse the fluorescence light is recorded proportional to $r(t)$. The detailed calculation of $L(t)$ quickly becomes tedious, as it requires explicit knowledge about the bath and the related coupling strengths. 
However it can often be modeled effectively as  $L(t)=\mathrm{exp}\left[-\left(t/T_2^*\right)^2\right]$ \cite{Hall14}. 
 
Since we are dealing with a pure dephasing dynamics, all common definitions of (non-)Markovianity coincide \cite{Zeng2011}. Explicitly, the dynamics is non-Markovian if and only if $\mathrm{d}{r}(t)/\mathrm{d}t >0$ for some time $t\geq 0$. On the other hand, the different ways to quantify the degree of non-Markovianity are not equivalent \cite{Vacchini2011,Addis2014}. In particular, we choose to measure the amount of non-Markovianity via the trace distance \cite{Breuer2009} which identifies non-Markovian evolutions as those with a back flow of information from the environment. 
By taking an integral over all the time intervals where the trace distance increases and maximizing over the couple of initial states, one can then define a measure of non-Markovianity $\mathcal{N}$. For the model at hand, this is simply given by
\begin{equation}
\mathcal{N} = \sum_{m} r(\tau^\prime_m)-r(\tau_m),
\label{eq:NMmeasure}
\end{equation}
where $m$ labels all intervals $(\tau_m,\tau^\prime_m)$ with $r(\tau^\prime_m)-r(\tau_m) > 0$. Indeed, we have $\mathcal{N}=0$ for a Markovian evolution, corresponding to a monotonic decay of the electronic coherence, while any revival in the coherence will induce an increase of the non-Markovianity. 

In order to analyze the collected data and predict unperformed measurements we set up a probabilistic model (see also the Supplementary Material \cite{SM}). Given a prior (probability) distribution $\mathcal{P}(\Theta)$ on a set $\Theta$ of paramaters to be estimated, Bayes theorem provides the posterior distribution $\mathcal{P}(\Theta \vert \mathcal{X})$ quantifying the probability that the model employing $\Theta$ accurately describes the data $\mathcal{X}$, $\mathcal{P}(\Theta | \mathcal{X}) \propto \mathcal{P}\left(\mathcal{X} \vert \Theta\right) \mathcal{P}(\Theta)$.
Here $\mathcal{P}\left(\mathcal{X} \vert \Theta\right)$ is the likelihood that we obtain $\mathcal{X}$ given $\Theta$. A Markov Chain Monte Carlo (MCMC) algorithm samples the posterior distribution after specifying likelihood and prior yielding two main advantages: First, any correlation between different parameters is inherent to the model, and second, error bounds arise as a natural result from the sampling process. Using probability theory, marginals for all elements in $\Theta$ can be obtained \cite{Kruschke2015, Sharma2017}. 
\\

\textit{FID decay under the influence of the bath.---}
\begin{figure}[t!]
\begin{centering}
\vspace{0cm}
 \hspace{-0.3cm} \includegraphics[width=1\columnwidth]{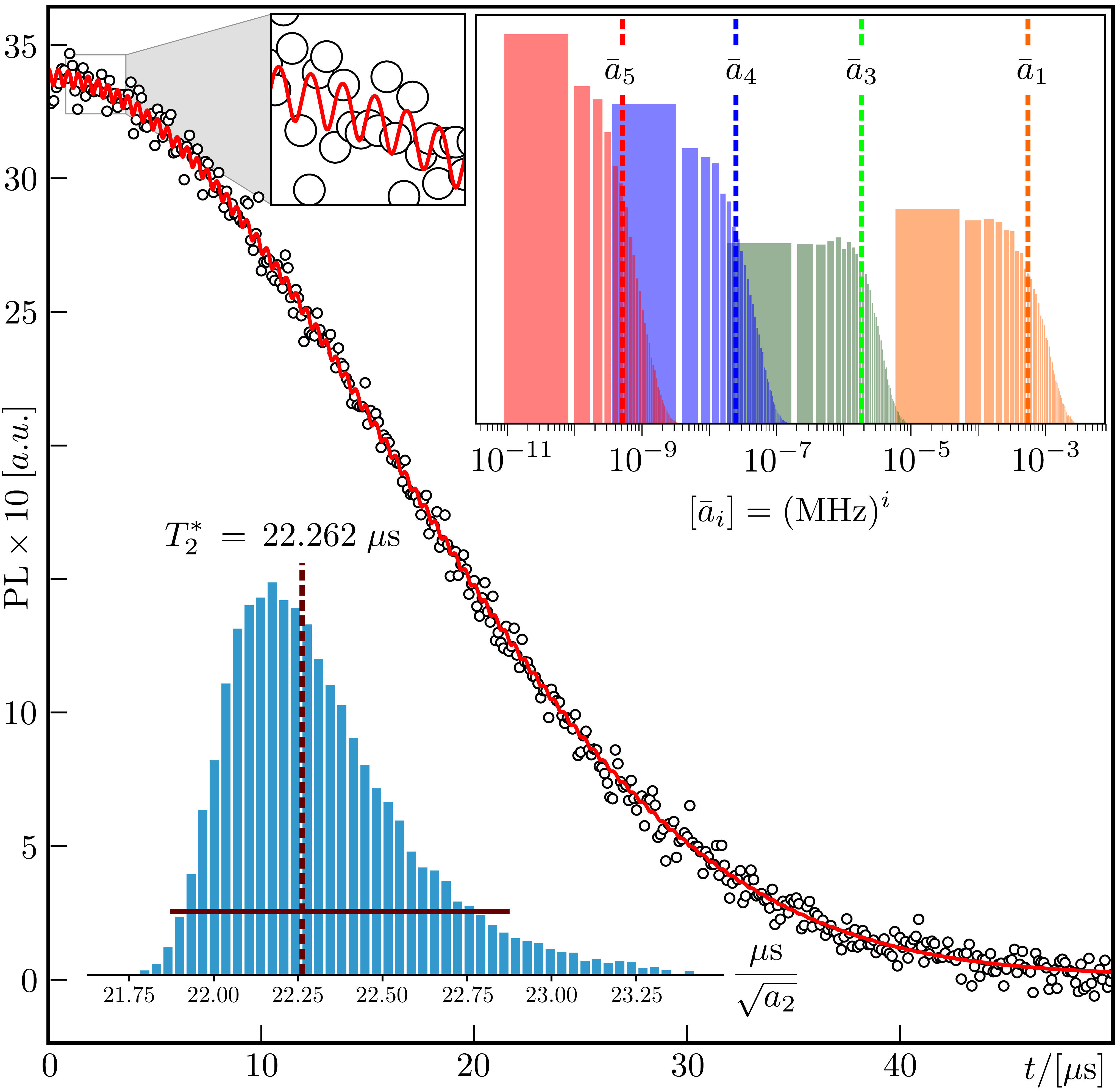} 
\par\end{centering}
\caption{(Color online) FID of curve of the NV center in units of the measured photoluminescence (PL). The negligible values of the decay constants $\bar{a}_i$ in the top right histogram of sampled values supports the purely Gaussian shape of $L(t)$. The histogram in the lower left assembles the distribution for $T_2^*$ with the HPD interval marked by the horizontal line. From the initial oscillations at short times, the Bayesian method can extract also other parameters (distributions not shown), as $p=0.972$ with HPD $[0.943,\, 1]$, $\phi = 0.191$ with HPD $[0.151,\,0.227]$ and $A_{\parallel}=2\pi\cdot2.143\,\mathrm{MHz}$ where HPD $2\pi\cdot[2.137,\,2.148]\,\mathrm{MHz}$.}   
\label{fig:FID} 
\end{figure}
In a preliminary experiment we explore the agreement of the FID envelope induced by $H_R$ with a monotonic decay to exclude contributions to a non-Markovian evolution. Therefore, polarization of the \Nitro spin is performed such that $p_0=1$ (and $R(\phi)\equiv\mathds{1}$). This enables a measurement of $|L(t)|$. Fig.\ref{fig:FID}(a) shows the FID envelope. We model the observed likelihood distribution by a normal distribution with a mean $\mu  = r(t)+d$ and $|L(t)| = \mathrm{exp}\left(-\sum_{i=0}^5 a_i t^i \right)$, see also Eq.\eqref{eq:BlochLength}. Here, $a_0$ is a constant to normalize the measured contrast and $d$ a possible bias in the asymptotic regime. After 50000 iterations of the chosen sampling algorithm \cite{SM}, we plot the red curve using the medians of the sampled parameters and the marginals of the posterior distribution for all $a_{i>0}$ in the insets. The experimentally measured contrast at specific times is shown with black dots. The FID envelope is well characterized by a $L(t)=\mathrm{exp}\left[-\left(t/T_2^*\right)^2\right]$, i.e. the dynamic is fully Markovian. We extract the characteristic timescale from the marginal of $a_2$ (we take the median as the point estimate and denote it by $\bar{\bullet}$) and obtain $T_2^*=22.262\,\mu s$ where the $95\%$ highest posterior density (HPD) interval (i.e. $95\%$ of sampling values lie in that region) is $[21.878, 22.868]\, \mu s$. Coherence envelopes of this form are extremely useful for frequency estimation using entangled states, since the Gaussian decay ensures a super-classical scaling of the estimation error with the number of probes \cite{Chin2012}. 

A careful examination of the short time regime reveals oscillations in the FID curve (see inset in Fig.\ref{fig:FID}), suggesting that the nitrogen spin is not fully polarized, as confirmed by the Bayesian method exploiting Eq.\eqref{eq:BlochLength} of our model \cite{SM}. The procedure is able to extract the different contributions to the decay stemming from the bath ($L(t)$), but also the parameters describing the \Nitro spin, i.e. we obtain the coupling strength $A_{\parallel}$ and the parameters $\phi,\,p$ (values see Fig.\ref{fig:FID}) for the population distribution [up to the symmetry in $\ket{m_I=\pm1}$ which is not resolvable in such experiment, see Eq.\eqref{eq:BlochLength}].
\\

\textit{Tuneable non-Markovianity.---}
\begin{figure}[t!]
\begin{centering}
\vspace{0cm}
 \hspace{-0.3cm} \includegraphics[width=1\columnwidth]{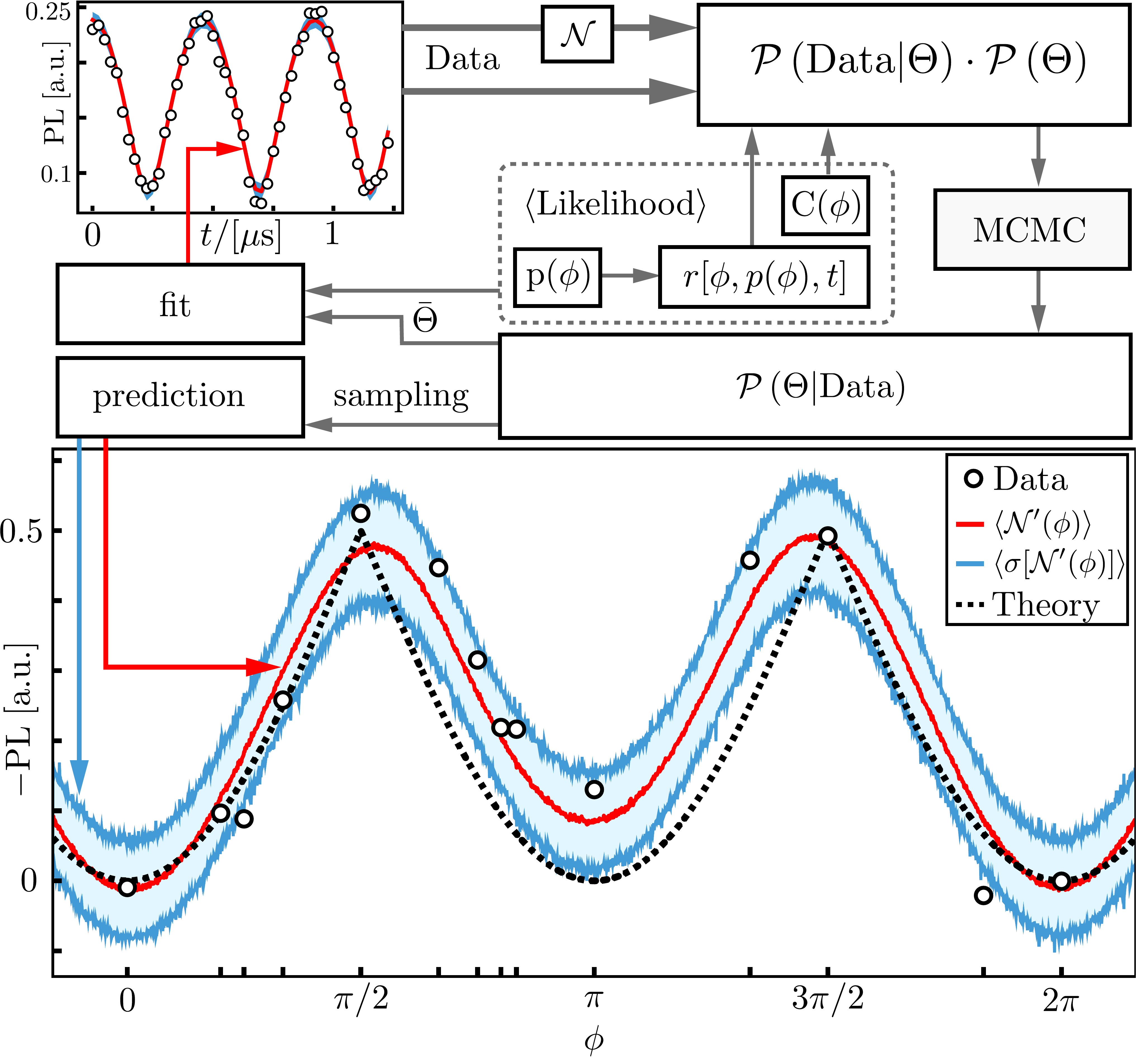} 
\par\end{centering}
\caption{(Color online) Ramsey measurements are performed for different values of $\phi$ (top left, only $\phi=2\pi/3$ is shown, black circles) and the non-Markovianity measure $\mathcal{N}^\prime$ (lower plot, black circles) is evaluated, which are both fed as observations into the likelihood distribution. The expectation value of the likelihood is constructed according to $\mathcal{N}^\prime$, while the prior distributions are taken as normal distributions around physically reasonable values \cite{SM}. The HPD parameter set $\bar{\Theta}$ can be plugged into the model defining the likelihood, which results in the maximum a posteriori inference (red curve) to the Ramsey data. The posterior distribution is sampled for different and, crucially, not measured values of $\phi$. This results in an expectation value $\langle \mathcal{N}^\prime \rangle$ which is taken with respect to the posterior, shown as the red curve in the lower plot along with the blue region marking the standard deviation. The black dotted line corresponds to the theory result neglecting the varying readout contrast and remaining population in $\ket{m_I=-1}$.}
\label{fig:NM_measure} 
\end{figure}
An imperfectly polarized \Nitro spin, i.e. a coherent or incoherent mixture of $I_z$ eigenstates, induces oscillations on the electron spin coherence (see Fig.\ref{fig:FID}), consequently the reduced electron spin state undergoes a non-Markovian evolution. Vice versa, any population of the \Nitro spin state undergoes the conditional evolution governed by the Hamiltonian \eqref{eq:rotHam}. At the point of maximal achievable correlations [Fig.\ref{fig:FigTheory}(b), right], the reduced state of the electron spin has reached its point of minimal coherence. Following is an increase in coherence corresponding to a reduction of correlations \cite{NOTE1} between the two spins. Consequently, changing the orientation of the polarization of the nitrogen spin allows to control the non-Markovianity of the electron spin in a continuous manner. 

In order to measure experimentally the amount of non-Markovianity, we follow again the Ramsey scheme, Fig.\ref{fig:FigTheory}(a). After polarization, the nitrogen spin population can be manipulated by a resonant radio-frequency pulse $R(\phi)$ to create the nuclear spin state $\ket{\psi_I}= \sin\left(\phi/2\right)\ket{m_I=0} + \cos \left(\phi/2\right)\ket{m_I=1}$. We track the evolution of the electron spin for 14 different values of $\phi$ up to a maximum time of $T=1.226\,\mu\mathrm{s}$ and record the oscillations in the coherence. 

Let us now describe the probabilistic model for this specific setup (see \cite{SM} for further details). First, note that a theoretical measure of non-Markovianity as defined in Eq.\eqref{eq:NMmeasure} requires processed data (e.g. fits). Otherwise, fluctuations will dominate the measure, e.g. for a constant coherence function fluctuations of the measurement results accumulate and give a positive measure. To avoid this issue we exploit the oscillatory nature of the modulation and stop the recording of the oscillation before finishing an integer number of periods. The requirement of an increase of the coherence in Eq.\eqref{fig:NM_measure} is then relaxed and the sum runs over all intervals, so that the fluctuations in the data are averaged out. The model for the measure then possesses the simple form $\mathcal{N}^\prime(\phi) = C(\phi) \lbrace r[\phi,p(\phi),T]-1\rbrace$ where $C(\phi)$ is a parameter describing the measurement contrast \cite{Steiner10} and $1-p(\phi)$ the population left in $\ket{m_I=-1}$. We infer the model on the measured data to obtain the information of these functional dependencies, see the upper part of Fig.\ref{fig:NM_measure}. Afterwards, the obtained posterior distribution is used to predict $\mathcal{N}^\prime(\phi)$ for different values of $\phi$ by drawing multiple samples and calculating the mean values. 

The theoretical and experimental results are reported in Fig.\ref{fig:NM_measure}. In the lower part, black dots mark $\mathcal{N}^\prime(\phi)$ for the 14 measured instances of $\phi$. The theory curve according to Eq.\eqref{eq:NMmeasure} in black (dotted) is rescaled to match the values of the contrast. Its deviation from the red curve, which illustrates the expectation value of $\mathcal{N}^\prime(\phi)$ sampled with respect to the posterior distribution, is due to the fact that the Bayesian model includes the angle dependent contrast and the nitrogen population left in $\ket{m_I=-1}$. In other words, the posterior distribution predictions of our parameters, together with the model in Eq.\eqref{eq:BlochLength} enable us to simulate further measurements of the experiment. We show the standard deviation of the sampling as the blue region, which covers most of the actual measurements. This standard deviation is due to error sources not included explicitly in the model, e.g., remaining population of the electron spin in $\ket{m_s=1}$ or drifts in the experimental setup.
\\

\textit{Conclusion.---} We experimentally demonstrated the control of the degree of non-Markovianity in the dynamics of an NV electron spin. To that end, we first examined the FID envelope and employed a Bayesian probabilistic model to ensure that the degree of non-Markovianity is induced by the residual background resulting mainly from a nuclear spin environment. Subsequently, we exploited the inherent \Nitro spin to induce modulations on the electron spin coherence. The \Nitro provides us with a natural source of non-Markovianity, which, depending on its initial preparation, will be able to exchange a certain amount of information with the electron spin, influencing the evolution of the latter. Despite of the initial control the \Nitro remains a natural source of non-Markovianity as no further interventions after the preparation have to be performed. The experimental effort is kept sufficiently low by using Bayesian techniques, which allow to predict the shape of the considered non-Markovianity measure. Let us also mention that the scheme presented may be extended by the utilization of strongly coupled \Carb spins or interacting NV centers. Using the same technique as described here, additional parameters to shape the evolution can be introduced. Further modifications could be implemented as well via a classical driving with random, but temporally correlated amplitude.

In summary, the configuration investigated here allows the assembly of an experimental platform with intrinsic non-Markovianity. This provides a building block for the systematic investigation of memory effects in the performance of, e.g., quantum sensors and quantum metrology protocols, as well as facilitating the controllable inclusion of memory in quantum simulations of open quantum system dynamics.

Note added: During the writing up of our results, related experimental
results on non Markovian features of NV center dynamics were reported
in \cite{Wang2018} and \cite{Peng2018}.

\textit{Acknowledgements.---}
This work was supported by the ERC Synergy grant BioQ, the EU project QUCHIP, the DFG, BMBF and the Volkswagenstiftung.
We acknowledge discussions with the team of Jiangfeng Du. We thank Matthew Markham for sample preparation. 

\begin{appendix}
\renewcommand\thefigure{\thesection.\arabic{figure}}
\setcounter{figure}{0}

\end{appendix}

\pagebreak{}\clearpage{}

\onecolumngrid

\begin{center}
\textbf{\large{{{Supplemental Material:\\ Controllable Non-Markovianity for a Spin Qubit in Diamond}}}}{\large{{ }}} 
\par\end{center}

\twocolumngrid

\setcounter{equation}{0} \setcounter{figure}{0} \setcounter{table}{0}
\setcounter{page}{1} \makeatletter \global\long\def\theequation{S\arabic{equation}}
 \global\long\def\thefigure{S\arabic{figure}}
 \global\long\def\bibnumfmt#1{[S#1]}
 \global\long\def\citenumfont#1{S#1}

\section{Experimental Details}

We present an introduction to the experimental platform, namely the nitrogen-vacancy center and the setup used to perform the measurement.

\subsection{The nitrogen vacancy center}

\begin{figure}[t!]

\begin{minipage}{.5\linewidth}
\centering
\subfloat[]{\label{fig:nv}\includegraphics[scale=.25]{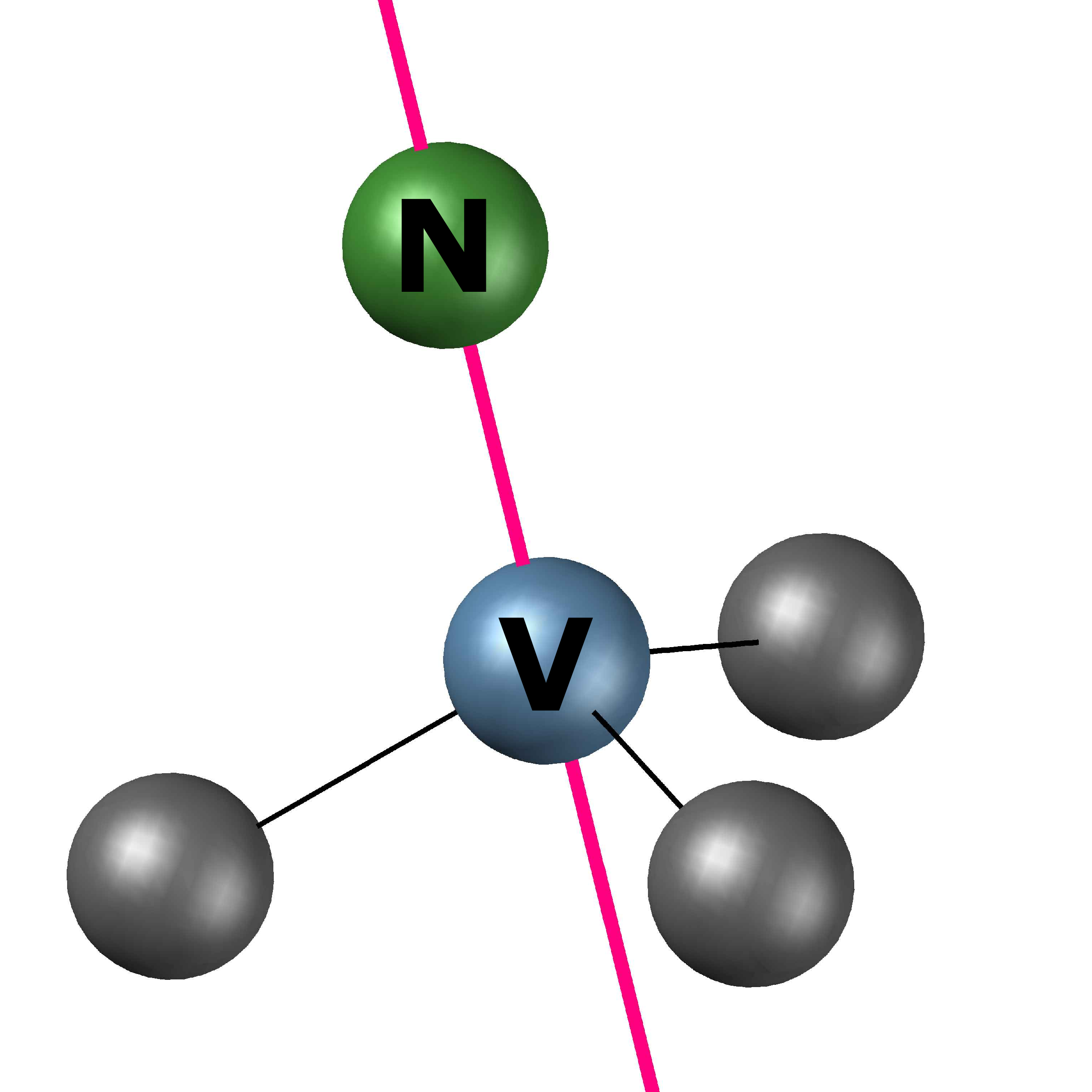}}
\end{minipage}%
\begin{minipage}{.5\linewidth}
\centering
\subfloat[]{\label{fig:splitting}
\begin{tikzpicture}[scale=1]
\tikzset{
	level/.style   = { ultra thick, black },
	connect/.style = { dashed, black }
}
\draw[level] (0,0) -- (1,0);
\draw[level] (0,0.25) -- (1,0.25);
\draw[level] (0,0.35) -- (1,0.35);

\draw[level] (0,3) -- (1,3);
\draw[level] (0,3.25) -- (1,3.25);
\draw[level] (0,3.35) -- (1,3.35);

\draw[ultra thick, ->, green] (0.25,0) -- (0.25,3.5);
\draw[ultra thick, ->, red] (0.5,3.5) -- (0.5,0);

\draw[dotted] (2.25,0.35) circle (0.5);

\draw[level] (2,0.35) -- (2.5,0.35);
\draw[level] (2,0.15) -- (2.5,0.15);
\draw[level] (2,0.55) -- (2.5,0.55);

\draw[level] (1.6,1.6) -- (2.6,1.6);
\draw[thick, ->] (0.72,3.35) -- (2.1,1.6);
\draw[thick, ->] (2.1,1.6) -- (0.8,0);

\draw[connect] (1,0.35) -- (2,0.35);
\draw[connect] (1,0.35) -- (2,0.15);
\draw[connect] (1,0.35) -- (2,0.55);

\node (a) at (1.4,1.8) {\textbf{${}^{1}\text{A}$}};
\node (a) at (-0.25,0.6) {\textbf{${}^{3}\text{A}$}};
\node (a) at (-0.25,3.6) {\textbf{${}^{3}\text{E}$}};

\end{tikzpicture}
}
\end{minipage}\par\medskip

\caption{Properties of the nitrogen-vacency center. Panel (a) illustrates the geometry of the NV center. It exhibits a symmetry for rotations of $2\pi/3$ around the axis connecting the nitrogen atom and the vacency. The corresponding simplified energy-level scheme is shown in (b). The ${}^3\text{A}$ electronic spin ground state triplet is excited by a green laser (green arrow) and subsequently decays according to the electronic spin state. Spin preserving radiative decay is indicated by the red arrow. The $\ket{m_S=\pm1}$ states preferentially decay by a non-radiative path (grey arrows) through an intermediate singlet state ${}^1\text{A}$ to the $\ket{m_S=0}$ state, which induces a difference in photoluminescence for the different inital states and enables polarization into $\ket{m_S=0}$. Additionally, the $\ket{m_S=\pm1}$ levels are split by the hyperfine interaction with the inherent nitrogen nuclear spin (three levels in the dotted circle shown exemplary for $\ket{m_S=-1}$).}
\label{fig:main}
\end{figure}

The nitrogen vacancy center (NV) is a point defect in the diamond lattice (Fig.\ref{fig:nv}), where it replaces two adjacent carbon atoms. It consists of a nitrogen at the first lattice site, while the other one remains empty. The three dangling carbon bounds donate three electrons to the NV, while the nitrogen atom possesses two free electrons. Together with an additional electron of an external donor, the NV center can form a negatively charged state which has an electronic spin of $S=1$. In the electronic ground state, this forms a spin triplet ${}^3\text{A}$ with a zero field splitting of $\Delta = 2\pi\cdot 2.87~\text{GHz}$ between the $\ket{m_{S}=0}$ and $\ket{m_{S}=\pm 1}$ states \cite{SManson06}. Hyperfine interaction with the inherent nitrogen nuclear spin results in further splitting of the $\ket{ m_{S}=\pm 1}$ states, depending on the nitrogen isotope. ($\text{A}_{\parallel}=2.14~\text{MHz}$ for the used ${}^{14}\text{N}$ isotope \cite{SFelton09}, $I=1$). This splitting is sketched in Fig.\ref{fig:splitting}.
\\

The preparation and readout of the electron spin is performed by optical excitation of the  ${}^3\text{A}$ state into the ${}^3\text{E}$ state. This transition is spin preserving, i.e. the population distribution in the $\ket{m_S}$ is not touched. The decay back to the ${}^3\text{A}$ is however strongly spin selective. The $\ket{m_S=0}$ state radiatively decays into its ground state, while the $\ket{m_S=\pm1}$ mainly passes through a non-radiative inter-system crossing to a metastable singlet state between excited and ground state \cite{SFelton09}, which also decays preferentially into the $\ket{m_S=0}$ ground state, i.e. this transition is not spin preserving. Firstly, this results in a higher intensity of the $\vert m_{S}=0\rangle$ state hence this difference in luminescence is used to determine the electron spin state. Therefore we will call the $\ket{m_S=0}$ state a  ``bright" state, while we refer the $\ket{m_S=\pm1}$ as ``dark" states. Secondly, long enough optical pumping polarizes the electron spin into the $\ket{m_S=0}$ ground state. 

The application of an external magnetic field along the symmetry axis of the NV center (i.e. an axis through the nitrogen atom and the vacency), splits the "darker" $\vert m_{S}=\pm 1\rangle$ states from each other due to the Zeeman effect. In this work, we employ the $\ket{ m_{S}=0}$ and $\ket{m_{S}=-1}$ states as our working transition for the artificial qubit.
Furthermore, the application of the external magnetic field close to $500\,$G (here we used $453\,$G) sets the steady state for optical pumping to $\ket{m_S=0,m_I=+1}$ \cite{SJacques09}. This is due to a spin level anticrossing in the ${}^3\text{E}$ state which allows energy conserving spin flip-flop processes between the electron and nitrogen spin. 

For an in depth description of the NV center and its properties we refer to reference \cite{SDoherty13}. 

\subsection{Sample information}

The diamond used in this work is a low nitrogen electronic grade diamond, grown by chemical vapor deposition with a depleted ${}^{13}\text{C}$ concentration of $0.2\,\%$ (natural concentration $1.1\,\%$). The used NV center is located deep (few $\mu$m) below the diamond surface. The external magnetic field had a strength of $453~\text{G}$.

A wire spanned over the diamond surface is used to realize coherent manipulations of the electron spin transition (microwave) or nitrogen spin transitions (radio-frequency). During each measurement, we refocus the position of the NV center every 40 seconds to overcome drifts in the optical setup. A precise measurement of the microwave transition frequency of the electron spin every 300 seconds ensures the elimination of possible transition frequency detunings during the experiment.

\subsection{FID measurement}
\label{sec:noise}

\begin{figure}[t!]
\centering
\begin{tikzpicture}[scale=1]
	\node [draw, fill=green!20, minimum height=0.8cm] (a) at (0.6,0.4) {pol.};
    \node [draw, fill=blue!20, minimum height=0.8cm, minimum width=1.4cm, right=0.2cm of a] (b) {$\left(\frac{\pi}{2}\right)_{\text{x}}^{\text{NV}}$};
    \node [draw, fill=gray!20, minimum height=0.8cm, right=0.2cm of b] (c) {$t$};
    \node [draw, fill=blue!20, minimum height=0.8cm, minimum width=1.4cm, right=0.2cm of c] (d) {$\left(\frac{\pi}{2}\right)_{\text{x/y}}^{\text{NV}}$};
    \node [draw, fill=green!20, minimum height=0.8cm, right=0.2cm of d] (e) {readout};
	\draw[ultra thick, ->] (0,0)  -- (6.7,0) node(xline)[right] {$t$};
\end{tikzpicture}
\caption{Pulse sequence for the free induction decay measurement. After the initial polarization a $\pi/2$ microwave pulse flips the electron spin to a superposition state in the equatorial plane, where it freely evolves for a given time $t$. Afterwards a second $\pi/2$ pulse flips the electron spin back to the z-axis (either bright or dark state), where the spin state is read out optically.}
\label{fig:noise_floor}
\end{figure}
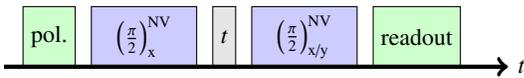

To examine the noise felt by the electron spin (besides the one originating the nitrogen spin), Ramsey experiments are performed to measure the free induction decay (FID) of the coherence. Therefore, we utilize the pulse sequence shown in Fig.\ref{fig:noise_floor}. After the initial polarization into $\ket{ m_{S}=0 }$, a $\pi/2$ microwave pulse flips the electron spin to a superposition state in the equatorial plane. During the free evolution time $t$, the electron spin picks up a phase, which originates from further impurities and spins in the diamond lattice. The origin of this phase can also be understood as a magnetic field along the $z$-axis of the electron spin, which fluctuates in a stochastic fashion. Afterwards, a second $\pi/2$ pulse flips the electron spin back to the $z$-axis. The experiment is performed twice, where either the dark or the bright state are chosen for the readout. The resulting dark and bright state fluorescence signals are subtracted to remove systematic errors in the measurement setup.

\subsection{Non-Markovianity control experiment}

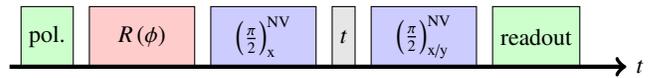
\begin{figure}[t!]
\begin{center}
\begin{tikzpicture}[scale=1]
	\node [draw, fill=green!20, minimum height=0.8cm] (a) at (0.2,0.4) {pol.};
	\node [draw, fill=red!20, minimum height=0.8cm, minimum width=1.4cm, right=0.2cm of a] (b) {$R\left(\phi\right)$};
    \node [draw, fill=blue!20, minimum height=0.8cm, minimum width=1.4cm, right=0.2cm of b] (c) {$\left(\frac{\pi}{2}\right)_{\text{x}}^{\text{NV}}$};
    \node [draw, fill=gray!20, minimum height=0.8cm, right=0.2cm of c] (d) {$t$};
    \node [draw, fill=blue!20, minimum height=0.8cm, minimum width=1.4cm, right=0.2cm of d] (e) {$\left(\frac{\pi}{2}\right)_{\text{x/y}}^{\text{NV}}$};
    \node [draw, fill=green!20, minimum height=0.8cm, right=0.2cm of e] (f) {readout};
	\draw[ultra thick, ->] (-0.3,0)  -- (7.9,0) node(xline)[right] {$t$};
\end{tikzpicture}
\caption{Pulse sequence to demonstrate of the precise control of coherence modulations via the nitrogen nuclear spin population. 
By applying a radio-frequency pulse $R(\phi)$ on the nitrogen nuclear spin, one can utilize the method described in Fig.\ref{fig:noise_floor} to measure the free induction decay. Since non-zero population in multiple nitrogen nuclear spin states result in a continuous alteration between a correlated and uncorrelated state, one will observe an oscillation in the coherence. Changing the phase of the last electronic $\pi/2$ pulse enables a full readout of its Bloch vector length.}  
\label{fig:nonmar_exp}
\end{center}
\end{figure}

We performed Ramsey experiments on the electron spin to demonstrate the precise control of coherence modulations via the population of the nitrogen nuclear spin. The pulse sequence, shown in Fig.\ref{fig:nonmar_exp}, follows the same procedure as the one, discussed in section \ref{sec:noise}. However, we extend the pulse sequence by a radio-frequency pulse to manipulate the population of the $\ket{m_{I}=1}$ and $\ket{m_{I}=0}$ states coherently after the initial polarisation ($\ket{m_{I}=-1}$ is not used in this work). The corresponding pulse length is determined by a Rabi measurement between the $\ket{m_{I}=0}$ and $\ket{m_{I}=1}$ states. Since shifts of the electrons transition (microwave-)frequency do not exceed $\sim 10~\text{kHz}$, the transition (radio-)frequency between the nitrogen nuclear spin states is assumed to be constant due to the smaller gyromagnetic ratio. Hence it is not refocused during the measurement. The coherent control enables us to prepare any arbitrary state $\ket{\psi} = \cos\left(\phi/2\right)\ket{m_{I}=1} + \sin\left(\phi/2\right)\ket{m_{I}=0}$ for the nitrogen nuclear spin.

After the radio-frequency pulse, the pulse sequence is identical to the sequence used in the FID experiment. The first $\pi/2$ microwave pulse flips the electron spin to the equatorial plane, where it freely evolves during the given time $t$. Afterwards the electron spin is flipped back to the $z$-axis to measure the fluorescence. Changing the phase of the last $\pi/2$ microwave pulse, i.e. either inducing a rotation around the $x$ of $y$ axis enables full readout of the $x$ and $y$ components of the Bloch vector which are required to calculate the length given in the main text.

\begin{figure}[t!]
\includegraphics[width=1\columnwidth]{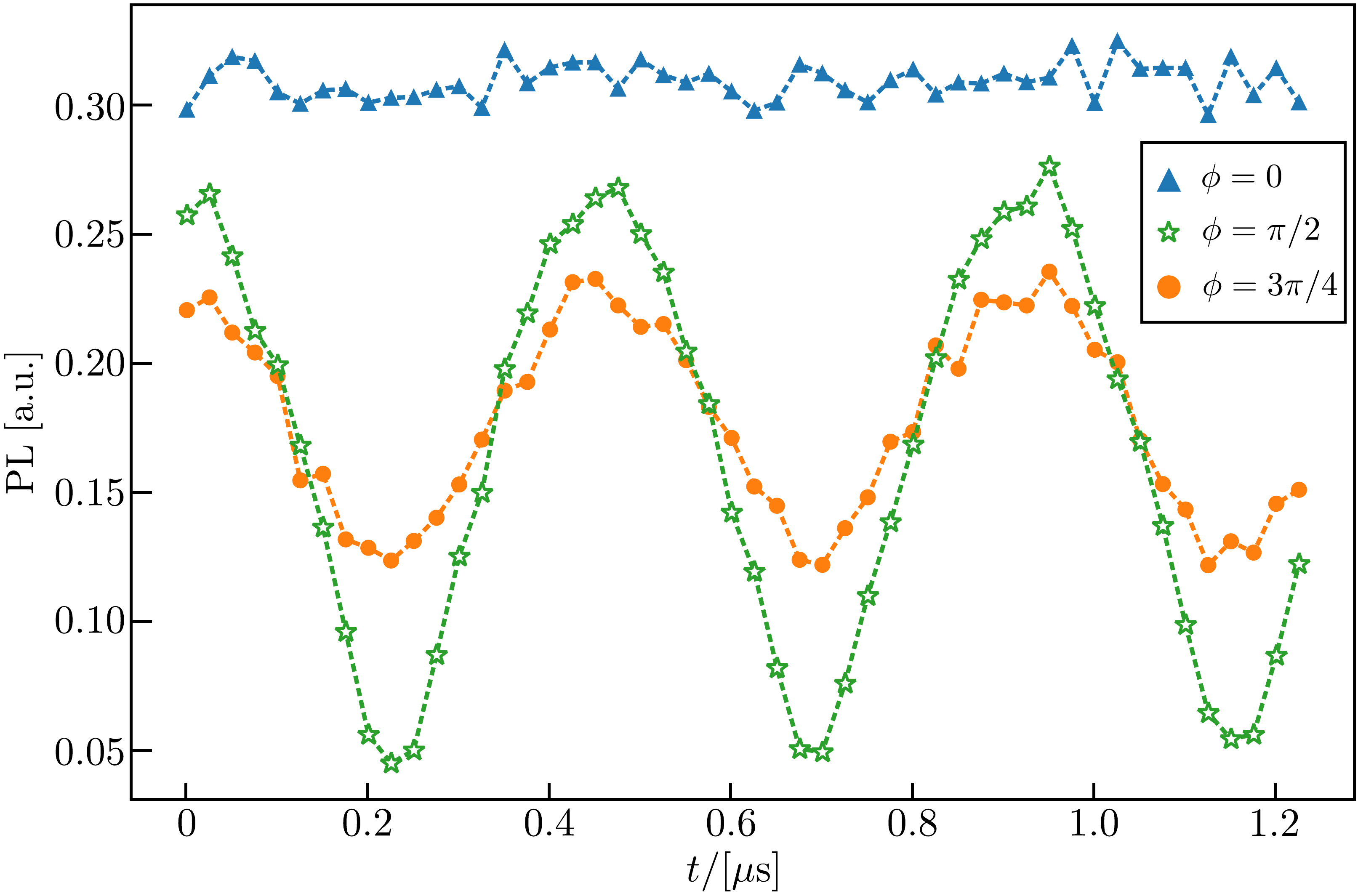}
\caption{Oscillations in the length of the electron spin's Bloch vector. The curves show three initial preparations of the nitrogen spin. Note the difference in amplitude and offset but equality in frequency and phase. The curve for $\phi=0$ (blue triangles) is supposed to show no oscillations theoretically, but of course it is influenced in practice by fluctuations in the measurement.}
\label{fig:oscillations}
\end{figure}

One observes an oscillating fluorescence signal, where three examples are shown in Fig.\ref{fig:oscillations}. The fluorescence signal corresponds to the absolute value of the Bloch vector length, which changes due to continuous correlation and decorrelation of the electron spin and nitrogen nuclear spin.

\section{Bayesian Inference}

We describe the basic formalism underlying the Bayesian approach adopted in the main text. 

\subsection{Introduction}

The problem of assigning observed outcomes to possible causes, is a daily faced problem in science. We may categorize possible causes with numbers which we call ``probabilities". Contrary to a \textit{frequentist approach}, where these probabilities are understood as the relative number of occurrences of precisely that cause (in the limit of infinite observations), the \textit{Bayesian approach} takes probabilities as a measure of certainty, i.e., ``how much one believes a certain cause to be the true one". Within the Bayesian approach, all causes, e.g. parameters to be inferred are probability distributions themselves, while the frequentist approach assumes them to be constant. For an in-depth review on Bayesian inference, see the references \cite{SSivia2006,SKruschke2015,SSharma2017}.

We denote a random variable for the parameters by $\T$ and a corresponding value by $\theta$. Note that in general $\theta$ can be a vector. The corresponding probability distribution is then written as $\p(\T)$. Let us introduce a second random variable for the measured data, $\X$, where the specific value is $x$. The probability, to make certain observation $\X=x$, i.e. a data point or a whole set of data, given some model $\T$ is then the conditional probability
\begin{equation}
\p(\X=x\vert\T=\theta) = \frac{\p(\X=x,\T=\theta)}{\p(\T=\theta)}.
\label{eq:Cond1}
\end{equation}
Note that the probability in the nominator on the r.h.s. denotes the probability to find $x$ \textit{and} $\theta$, while the l.h.s. is the probability to find $x$ \textit{given} $\theta$. From now on we will drop the notation like $\T = \theta$ and use the value alone. 

Instead of the probability to make a specific observation $x$, we are rather interested in the parameter $\theta$. Hence, we can formulate the conditional probability the other way round, i.e.
\begin{equation}
\p(\theta\vert x) = \frac{\p(x,\theta)}{\p(x)}.
\label{eq:Cond2}
\end{equation}
Usually, we do not have access to $\p(x,\theta)$, hence we combine Eq. \eqref{eq:Cond1} and \eqref{eq:Cond2} to obtain Bayes theorem,
\begin{equation}
\p(\theta\vert x)=\frac{\p(x\vert \theta) \p(\theta)}{\p(x)}.
\label{eq:Bayes}
\end{equation}
This theorem describes the desired object, i.e. the \textit{posterior probability distribution} $\p(\T\vert\X=x)$. This object quantifies ``how certain we are that a given $\T=\theta$ is the cause of the outcome $\X=x$". The r.h.s. of the theorem is specified in the following way. The \textit{likelihood function} for $\T$ is $\p(\X\vert \T)$. While it is a probability distribution for $\X$ given $\T$, we can also think of it as a function weighting the values of $\T$ to the usually fixed, since observed, values $\X =x$. The so called \textit{prior distribution} $\p(\T)$ is a powerful way to include prior knowledge of the parameter. For example, a flat distribution would correspond to no prior information. However, in the application described in the main text, e.g. for estimates of the coupling constant $A_\perp$, we may choose a Gaussian distribution with a mean value determined in earlier experiments. The only remaining quantity is the \textit{evidence} $\p(\X)$. While we are able to express $\p(\X) = \int \mathrm{d}\theta\, \p(\X\vert \T=\theta)\, \p(\T=\theta)$, we note that this quantity only serves as a normalization constant, hence we can neglect it and write Bayes theorem as 
\begin{equation}
\p(\T\vert\X) \propto \p(\X\vert\T) \, \p(\T)
\end{equation}
Therefore, the posterior distribution is always totally determined by the likelihood function and the prior distribution. \\

Possessing the posterior distribution allows the calculation of marginal probabilities, in case $\T$ has $n$ dimensions. The marginal distribution for one specific dimension of $\T$ quantifies the probability distribution for this dimension alone, irregardless of the distributions in other dimensions. In other words, if we have the parameter $\theta = (\theta_1, \theta_2, \dots ,\theta_n)$ then the marginal distribution for $\theta_i$ is given by
\begin{eqnarray}
\p(\theta_i\vert x) = \int_S \mathrm{d}\theta_1 \mathrm{d}\theta_2 \dots \mathrm{d}\theta_{i-1} \mathrm{d}\theta_{i+1} \dots \mathrm{d}\theta_n\; \p(\theta \vert x),
\end{eqnarray}
where $S$ the space of the allowed values for all $\theta_{j \not =  i}$. 

We can also use the posterior distribution to calculate a posterior predictive. Integrating over $\theta$ yields the \textit{posterior predictive distribution}
\begin{equation}
\p(\mathcal{Y}\vert \X=x) = \int_\T \mathrm{d}\theta\; \p(\T =\theta \vert \X = x)\, \p(\mathcal{Y}\vert \T = \theta) 
\label{eq:postPredDist}
\end{equation}
where $\mathcal{Y}$ are a second set of observations, which have not yet been detected in a real experiment. This is a powerful tool to first validate the obtained posterior distribution, but it can also be used to predict further observations due to the causes specified with the parameters in $\T$. 

Usually, the posterior distribution cannot be calculated analytically and even a numerical solution requires increasingly large effort, when the dimension of the distributions increases. However, one can use Markov-Chain-Monte-Carlo (MCMC) methods to sample the r.h.s. of Eq.\eqref{eq:Bayes} efficiently. This technical detail goes well beyond the scope of this work and many examples of these methods can be found \cite{SSharma2017, SKruschke2015}. Here we use the No-U-Turn Sampler \cite{SHoffman2014}, an extension to the Hamilton-Monte-Carlo MCMC algorithm \cite{SNeal1993}. The models in this work have been implemented using the PyMC3 software package for the Python programming language \cite{SSalvatier2016}.

Using these algorithms, the obtained posterior distribution is given in terms of the relative frequencies, with which a specific realization of $\T$ appears. Calculating the marginals, gives the relative frequencies of the values of a single parameter. The point estimate, which we denote by an overbar $\bar{\bullet}$, is then given by the median of the marginal posterior distribution (we neglect multimodal distributions in this work), however the uncertainty in this value is determined by the shape of the distribution. A natural way to summarize the distribution is the interval of highest posterior density (HPD). The HPD specifies the interval of values, which all have a higher probability than the values outside the HPD. Usually, the HPD is taken to cover a larger proportion of the distribution, e.g., as we also chose in the main text,  95\% of all values which occurred during the sampling. An easy analogy is the width of a standard derivation, which contains $95.45\%$ of its values within a region of width $4\sigma$ around its mean value. In case the marginal posterior distribution would be Gaussian, the HPD and the $4\sigma$ region would be equivalent.

\subsection{Free Induction Decay}
We recall the expression for the length of the Bloch vector, Eq.(4) in the main text:
\begin{eqnarray}
r(t) =  &&\Big[p_0^2+p_1^2+p_{-1}^2+2p_0(p_1+p_{-1})\cos(A_{\parallel}t)  \nonumber \\
&& +2p_1p_{-1}\cos(2A_{\parallel}t) \Big] ^{1/2}\,|L(t)|,
\label{eq:BlochLength_SM}
\end{eqnarray}
where our assumption for $L(t)$ is of the form
\begin{equation}
L(t)=e^{-\sum_{i=0}^5 a_i t^i }.
\end{equation}
As a first step, we define the parameters to be inferred. Note that we have $p_0 + p_1 + p_{-1} = 1$ and hence we can express these three parameters in terms of two as described in the main text, i.e. $p_1=p \cos^2(\phi/2)$, $p_0=p \sin^2(\phi/2)$ and $p_{-1}=1-p$. Therefore we define the parameters $\theta = (\lbrace a_i \rbrace, \phi, p, A_\parallel,\sigma)$. We are looking for the coefficients $\lbrace a_i \rbrace$ fixing the FID envelope, the populations of the nitrogen spin and the parallel coupling constant $A_\parallel$. To account for errors, we define $\sigma$ as the standard deviation of the measurements in the experiment. Each data point $j$ for the FID envelope in Fig.2 of the main text is uniquely determined by its time $t_j$ and its value, let's call it $x_j$. We construct the likelihood function in the following way. Each value $x_j$ of a random variable $\X = (\X_1,\dots,\X_n)$ is assumed to be a draw from a normal distribution of variance $\sigma^2$ and an expectation value $\mu_j = r(t_j)\vert_\theta$, i.e. we have
\begin{eqnarray}
\p(\X \vert \theta) \propto \exp\left\lbrace -\frac{[\X-r(T)\vert_{\theta}]^2}{2\sigma^2}\right\rbrace
\label{eq:normDist}
\end{eqnarray}
where we use the vector of measurement times $T=(t_1,\dots,t_n)$. Prior distributions for the parameters $\theta$ are also assumed either to be normally distributed around their expected value, e.g., $A_\parallel \sim N(2\pi\cdot 2.14\,\mathrm{MHz}, \sigma_{A_\parallel})$, or distributed according to a positive half normal distribution (all $a_{i}\geq 0$).
Note that the origin of $\sigma$ is not explicitly specified, but it is an inherent quantity of the model. As mentioned in the main text, it accounts for error sources not specified in the model, but an unnaturally large $\sigma$ may also indicate a falsely specified model.

\subsection{Non-Markovianity - Model}
\begin{figure}[t!]
\begin{centering}
\vspace{0cm}
 \hspace{-0.3cm} \includegraphics[width=1\columnwidth]{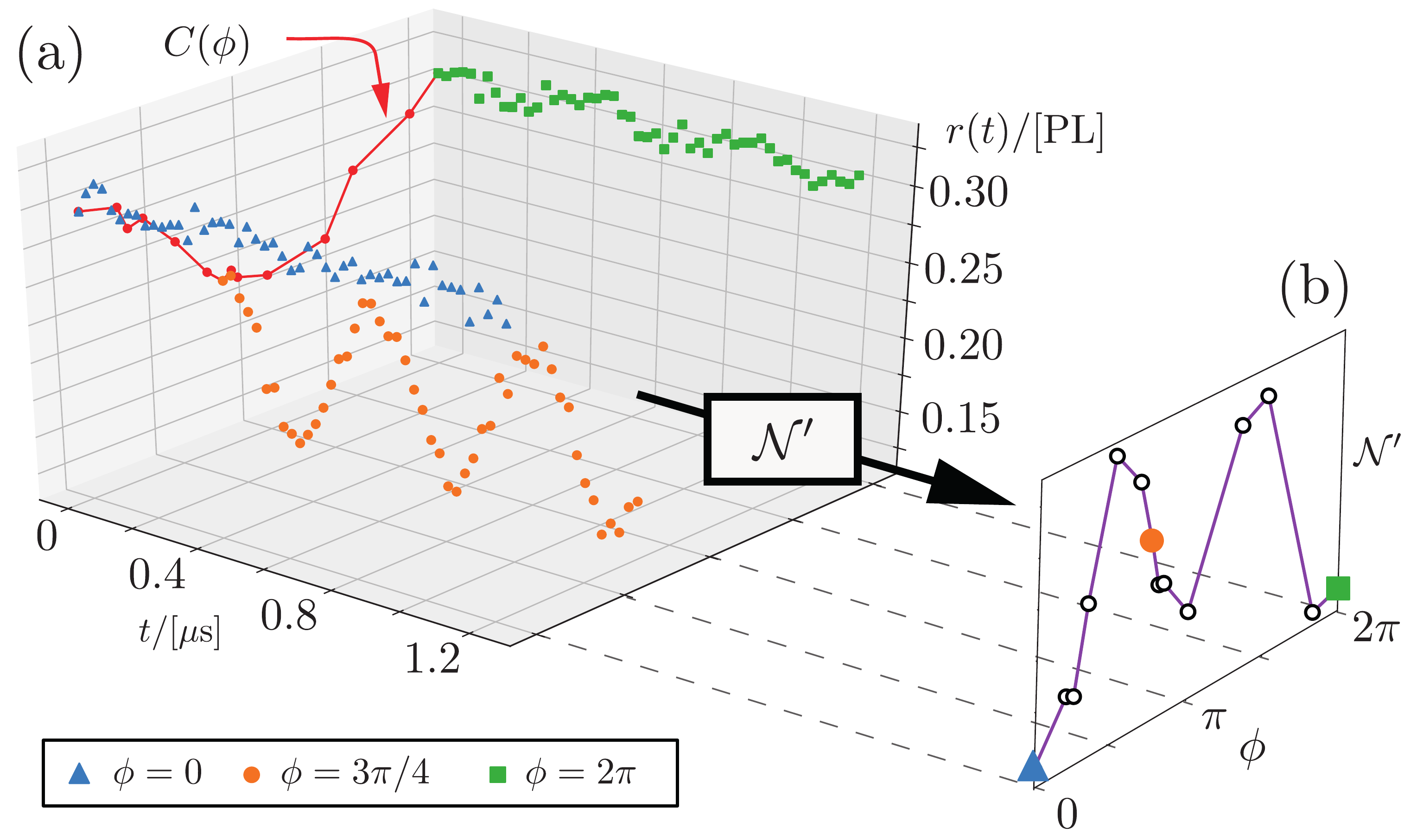} 
\par\end{centering}
\caption{Illustration of the collected data. For visibility in plot (a), we only show 3 exemplary data sets of the total 14 which have been collected. The angle of each set collected for the analysis is shown via a red dot at $t=0$, which also illustrates the measurement contrast depending on the polarization direction of the nitrogen spin. All curves have two full revivals at the same position as the one for $\phi=3\pi/4$ (orange dots). From this data, one can compute the modified measure of non-Markovianity $\mathcal{N}^\prime$ which results in panel (b). Each curve for a single instance of $\phi$ results in a single point of $\mathcal{N}^{\prime}$ (compare also Fig.3 in the main text).}   
\label{fig:NM} 
\end{figure}
Given the pure dephasing dynamics of the electron spin, it is enough to monitor its modulated coherence evolution as explained in the main text. 
To demonstrate controllable non-Markovianity, we monitored the length of the Bloch vector for 14 different initial preparations of the nitrogen spin, i.e. different values for $\phi$ (see the applied pulse angle of the radio frequency field). In the following, we want to illustrate the construction of the model used for the Bayesian inference.
\begin{enumerate}
\item We aim to describe the whole collected data by a common function, i.e. a model which gives the value of the coherence depending on the given point in time and the rotation angle of the nitrogen spin.   The Bloch vector is still described by Eq.\eqref{eq:BlochLength_SM}. Since we parametrized the population of the nitrogen spin, we have 
\begin{eqnarray}
r(t,\phi) &=&   \Bigg\lbrace 2(1-p)p \cos(2A_\parallel t) \cos^2\left(\frac{\phi}{2}\right)  \notag \\
&+&\frac{4-p(8-7p) + p^2 \cos(2\phi)}{4} \notag \\
&+&p \cos(A_{\parallel} t) [2-p+p\cos(\phi)] \sin^2\left(\frac{\phi}{2}\right) \Bigg\rbrace^{\frac{1}{2}}.
\end{eqnarray}
Since the maximal time of the free evolution was chosen such that $t\leq T=1.226\,\mu\mathrm{s}$, we assume $L(T)\approx L(0) \approx 1$ because of the long coherence time provided by the low $^{13}$C concentration.
\item Three of the 14 data sets are shown in Fig.\ref{fig:NM}(a). The red curve at $t=0$ (red circles mark the position of other data sets) illustrate the dependence of the readout contrast on $\phi$. To take this into account, we model the contrast via
\begin{equation}
C(\phi) = C_a\cos\left(C_\nu\,\phi\right)+C_b,
\end{equation} 
where $C_\beta$ are unknown constants to model amplitude $(a)$, frequency $(\nu)$ and offset $(b)$ of the modulation. Each of these are chosen as normally distributed variables.
\item Analogously to the contrast, we need to parametrize $p$ in terms of $\phi$. This is less straight forward, however a suitable parametrization can be found by mimicking Rabi oscillations found in driven three level systems in ladder configuration. We assume
\begin{equation}
p(\phi) = 1-\left[p_b+p_a\sin(p_\nu\,\phi+p_\varphi)\right].
\end{equation}
%
\item  Next, we need to define the measure of non-Markovianity. In Fig.\ref{fig:NM}, note the fluctuations for $\phi=0$ (blue triangles) and $\phi=2\pi$ (green squares). Ideally, these measurements would correspond to a constant which results in a zero value of the non-Markovianity measure. However, equally distributed fluctuations will be eliminated when we sum over all differences instead of only the positive ones. This results in the definition of the measure $\mathcal{N}^\prime$. Calculating the measure analytically, we arrive at
\begin{eqnarray}
\mathcal{N}^\prime(\phi) &=& C(\phi) \sum_{i=1}^{n-1} \left[r(t_{i+1},\phi)\vert_{p=p(\phi)} - r(t_i,\phi)\vert_{p=p(\phi)} \right] \notag \\ 
&=& C(\phi) \left[ r(T,\phi)\vert_{p=p(\phi)}-1\right],
\end{eqnarray}
which was already given in the main text (with $t_n = T$). The result for the measured data sets is also shown in Fig.\ref{fig:NM}(b). Note that we always have $r(T,\phi)\leq1$ so it is always $\mathcal{N}^{\prime}(\phi) \leq 0$. However, because of the periodicity of $r(t,\phi)$ in time (for all angles $\phi$, it is fixed by $A_\parallel$), this does not change the meaningfulness of the measure. In particular, because the induced oscillation has the same frequency for all $\phi$ and at $t=T$ all trajectories of $r(t,\phi)$ are in phase, the amplitude of the oscillation is sufficient to quantify the non-Markovianity of the evolution.
\item
The Bayesian inference model for the measure of non-Markovianity possesses a parallel structure. Crucially, our inferred parameters need to be fitted  to the modulations of the coherence, while they are at the same time required to describe the measure of non-Markovianity. That is, we have the parameters $\theta = (\lbrace C_\beta\rbrace, \lbrace p_\alpha \rbrace , A_\parallel)$ which need to hold for all observed coherence data $X$. However, we distinguish between the angle labels for the coherence $\phi$ and the non-Markovianity measure $\phi_{\mathcal{N}}$ for clarity. 
We rewrite Bayes theorem as
\begin{eqnarray}
\p(\theta, \sigma, \sigma_\mathcal{N}\vert X) &\propto& \p(X \vert \theta, \sigma, \sigma_\mathcal{N}) \; \p(\theta, \sigma, \sigma_\mathcal{N}) \notag \\
&=& \p_r(X_{\phi} \vert \theta, \sigma)\; \p_{\mathcal{N}}(X_{\phi_{\mathcal{N}}} \vert \theta, \sigma_{\mathcal{N}}) \notag \\ 
&& \times\; \p(\theta) \, \p(\sigma) \, \p(\sigma_\mathcal{N}),
\end{eqnarray}
where we could split the likelihood function these variables are conditionally independent in our probability model and mutually just depend on deterministically collected data. By $\sigma$ ($\sigma_\mathcal{N}$) we mark the standard deviation of the normal distribution used to model the likelihood function $\p_r$ ($\p_\mathcal{N}$). These distributions have the expectation values $r(t,\phi)\vert_{p(\phi)}$ and $\mathcal{N}^\prime(\phi_{\mathcal{N}})$ respectively [compare also Eq.\eqref{eq:normDist}]. 
\item We formulate the posterior predictive distribution according to Eq.\eqref{eq:postPredDist}.
\end{enumerate}

\subsection{Non-Markovianity - Results}
\begin{figure}[t!]
\begin{centering}
\vspace{0cm}
 \hspace{-0.3cm} \includegraphics[width=1\columnwidth]{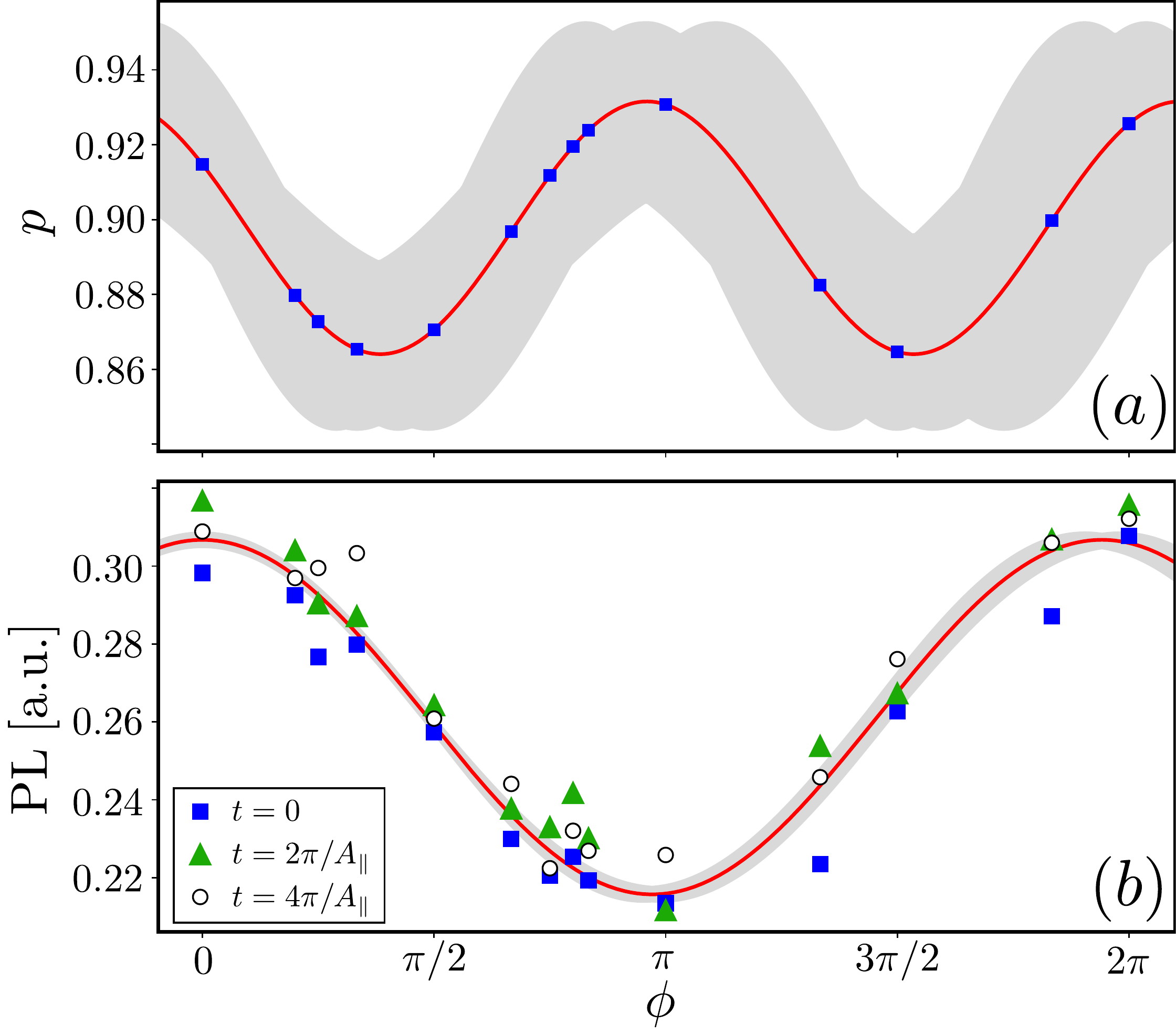} 
\par\end{centering}
\caption{Resulting models with parameters inferred by the Bayesian analysis. In (a) we plot the maximum posterior value (red curve) for the nitrogen population in the subspace $\ket{m_I = 0,1}$. The HPD interval is marked by the grey area and the blue squares indicate angles for which we measured the length of the Bloch vector. We emphasize that the blue squares do not represent a measured value of the population. However, in (b) we are able to compare our modeling of the contrast (red curve) with the contrast directly after initialization (blue squares) and the first two full revivals (green triangles, black circles). The HPD interval is set by the grey area.}   
\label{fig:NM_contrast} 
\end{figure}

We infer the posterior distribution for the model of the non-Markovianity measure as introduced above. The results are summarized in the following table, where the first two columns specify the properties of the prior normal distribution and the second two columns the point estimate (median of the marginal) and the HPD interval:
\begin{center}
\begin{tabular}{c||ccccc}
 & $\mu$ & $\sigma$ & point estimate  & HPD \\ 
\hline \hline
$C_a$ & 1 & 0.1 & 0.046 & $[0.044,\, 0.047]$  \\ 
$C_\nu$ & 0.3 & 0.1 & 1.030  &  $[1.011, \,1.050]$  \\ 
$C_b$ & 1 & 0.1 & 0.261  & $[0.260, \,0.262]$   \\ 
$p_a$ &0.02 & 0.01  & 0.034 & $[0.023, \,0.044]$   \\ 
$p_\nu$ &1.5 & 0.1 & 1.738 & $[1.611, \,1.858]$   \\ 
$p_b$ & 0.02 & 0.01 & 0.102 & $[0.091, \,0.112]$   \\ 
$p_\varphi$ &0 & 0.3 & -0.528 & $[-0.904, \,-0.134]$   \\ 
$A_\parallel\, /[\mathrm{rad}/\mu\mathrm{s}]$ & $4.2\pi$ & 0.5 & $2\pi\,2.169$  & $2\pi\, [2.165,\, 2.173]$   \\
$\sigma_{\mathcal{N}}$ & 0 & 1 & 0.060 & $[0.037, \,0.096]$   \\ 
\end{tabular} 
\end{center}
The value for the $\sigma$ values describing the standard deviations of each coherence function is not explicitly shown here, but the maximum posterior value for the largest of these is $0.018$. \\

In Fig.\ref{fig:NM_contrast} we illustrate the dependence of the population $p(\phi)$ and the contrast $C(\phi)$ on the polarization direction of the nitrogen spin. 
The plot in panel (a) shows the amount of population in the desired subspace of $\ket{m_I = 0,1}$. We remark again that the analytic solutions cannot distinguish between $\ket{m_I=\pm1}$. We assign the finite offset from unity to an imperfect polarization (i.e. $<100\%$) of the nitrogen before the radio-frequency pulse. At $\phi=0$ we can get a hint of the efficiency of the polarization: The red curves shows the maximum posterior value $\bar{p}(0)=0.915$, while the HPD interval (grey area) is fixed by $[0.891\; 0.943]$. During the pulse, population leaks to the $\ket{m_I=-1}$ state due to a non vanishing Rabi frequency between $\ket{m_I=0}$ and $\ket{m_I = -1}$, which is leading to the shown curve. \\
The change of contrast is plotted in Fig.\ref{fig:NM_contrast}(b). Along with the inferred curve, we also plot the coherence data at $t=0$ (blue squares, corresponds to the red curve in Fig.\ref{fig:NM}(a)) and the first two full revivals (green triangles, black circles), i.e. $r(t_k,\phi)=1$,  which occur at 
\begin{equation}
t_k = k\frac{2\pi}{A_\parallel}, \quad k \, \epsilon \, \mathds{N}.
\end{equation}
These times correspond to a totally decorrelated product state (see main text), i.e. the state of the electron spin is equivalent at all these points. Therefore the contrast is only determined by the polarization direction of the nitrogen spin which enables the comparison with our modeling of the contrast without calculating the impact of correlations. We remark, that from this plot we can again confirm that the assumption of a decoherence free evolution ($L(t)\approx1$) is justified. Otherwise, the coherence values of later times would strictly show less contrast than the ones at earlier times, which is not the case here. In particular, this is due to the long coherence time of the sample and the Gaussian shape of the envelope.


\begin{thebibliography}{10}
\bibitem{NOTE1} For an incoherent mixture of the nitrogen spin, the correlation would be classical. However, the induced effect [Eq.\eqref{eq:BlochLength}] on the reduced state of the electron spin would be the same.
\end{thebibliography}
\end{document}